# Next-level composite fermions

*A rich pattern of fractional quantum Hall states in graphene double layers can be naturally explained in terms of two-component composite fermions carrying both intra- and inter-layer vortices.*

Gábor A. Csáthy
Department of Physics and Astronomy, Purdue University, West Lafayette, IN, USA   gcsathy@purdue.edu;

and

Jainendra K. Jain
Department of Physics, The Pennsylvania State University, University Park, PA, USA   jkj2@psu.edu;

At first blush, accounting for the cornucopia of fractional quantum Hall states arising from the correlated motion of electrons in a single two-dimensional layer of electrons appears to be a difficult task. Luckily, it admits an elegant description in terms of emergent particles called composite fermions, which are born when each electron captures an even number of quantized vortices of the microscopic wave function. A large majority of the observed fractional quantum Hall states are accounted for as the integer quantum Hall states of composite fermions. Now, two groups writing in Nature Physics show that even more intricate composite fermions can exist in double layer structures [1,2].

The physics of electron correlations is significantly enriched when two layers are brought into close proximity. In such double layer systems the strength of the inter-layer correlation is a new degree of freedom that can be controlled by the separation between the layers. Fractional quantum Hall states are allowed that are not present in a single layer. The study of double layers of GaAs/AlGaAs provided early examples of inter-layer correlated fractional quantum Hall states. One such example is the fractional quantum Hall state forming at $\nu = 1/2$, a state associated with a Bose-Einstein condensation of interlayer excitons, formed when each electron binds a hole in the opposite layer [3]. (Here $\nu$ is the filling factor associated with one of the layers, defined as the number of electrons per vortex.) Another inter-layer fractional quantum Hall state in GaAs/AlGaAs forms at $\nu = 1/4$ [4,5], which is identified with the "331" wave function proposed by Halperin [6].

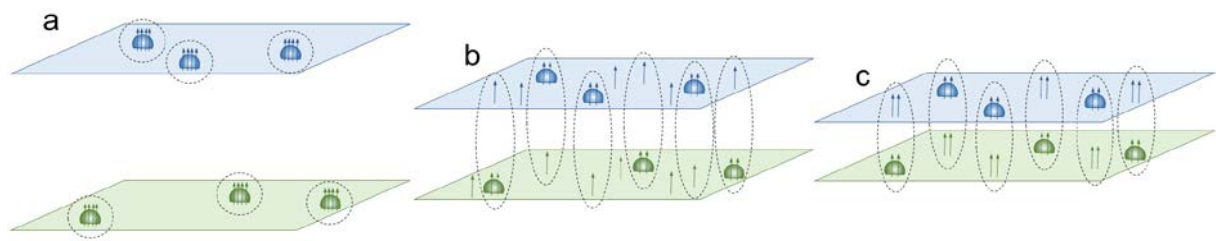

Figure 1. **Evolution of electronic correlations as a function of layer separation at $\nu = 1/4$**. Electrons are depicted as the round objects, while quantized vortices as arrows. Composite fermions are encircled by dashed curves. Composite fermions in panel a are single component particles as they capture only intra-layer vortices. In contrast, composite fermions at smaller layer separations are two-component, since they involve both intra- and inter-layer vortices. Two-component composite fermions of different flavors are shown in panels b and c. Panel b produces a fractional quantum Hall state, because the composite fermions in each layer are at filling factor $\nu^{CF} = 1$ (one composite fermion per unattached vortex). Both side panels represent Fermi seas of composite fermions experiencing zero effective magnetic field (no unattached vortices).

The possibility of a larger class of fractional quantum Hall states was revealed by a generalization of the composite fermion theory to double layer systems in which electrons bind both intra- and inter-layer vortices, thereby generating "two-component" composite fermions [7]. Fig.1 depicts three such constructions for the state at $\nu = 1/4$, which has four vortices per electron in each layer. In Fig.1a, where the layers are far apart, each electron captures four intra-layer vortices, leaving no room for inter-layer correlations. As the two layers are brought closer, composite fermions carrying two intra-layer vortices combine first with one vortex and then with two vortices in the opposite layer (Figs. 1b and 1c). Because composite fermions only "see" the unattached vortices, they form Fermi seas in panels a and c, and an integer quantum Hall state in panel b. Similar constructions are possible at other filling factors. Despite theoretical expectation, the existence of these more intricate composite fermions and their quantum Hall states had not been demonstrated in experiments.

Now, Liu and colleagues [1] and Li and colleagues [2], have taken a significant step toward realizing this plethora of inter-layer correlated states of two-component composite fermions, while also revealing surprising behaviour not anticipated by theory. These observations have become possible only because of several innovations in the layering of the graphene based structures, which have resulted in a substantial improvement of the quality of electron transport measurements. First, the insertion of a boron nitride layer for suppressing direct tunnelling between the graphene layers also effectively eliminates interfacial impurities (such impurities plague barriers of high aluminum content in GaAs/AlGaAs double well structures). Second, the authors take advantage of improvements afforded by the use of graphite gates, as opposed to the usual metallic ones. Additional improvement specific to transport is achieved by employing the Corbino geometry that avoids scattering near the physical edge of the graphene sample. The use of graphene double layers has also enabled access to a larger range of parameters, including previously unexplored region of very small layer separations.

Coulomb drag experiments -where current is driven through one layer and the response is measured in the other layer - and measurements in imbalanced densities allowed by the dual gated samples were instrumental in establishing the existence and nature of interlayer correlations [1,2]. In addition, while the Corbino geometry has not allowed access to drag measurements, it revealed an impressive array of incompressible states in double layer systems [2], rivalling the richness of fractional quantum Hall effect seen in single layers. Most of these are in excellent agreement with the pattern expected from two component composite fermions. For example, the measured values of the Hall resistances in the drag and drive layers at filling factor ν = 2/5 are consistent with the formation of a specific type of inter-layer correlations embodied by the composite fermions with 1 interlayer and 2 intra-layer vortices (as shown in Fig.1b).

These experiments also show incompressible states that do not correspond to integer fillings of composite fermions. For example, both papers report a fractional quantum Hall state at ν = 3/7 [1,2]. This would be unremarkable for uncorrelated layers, because while each electron sees 7/3 vortices, each composite fermion sees 7/3-2=1/3 vortices, giving an effective filling of composite fermions $\nu^{CF}$ = 3. However, drag and drive experiments show that composite fermions also bind one interlayer vortex, which gives $\nu^{CF}$ = 3/2. Naively, a fractional quantum Hall state is not expected at this filling factor. One may speculate that a gap opens here and at certain other half-integer values of $\nu^{CF}$ because of further correlations between composite fermions resulting in either inter-layer exciton condensation or Cooper-like pairing. A quantitative account of this physics remains a challenge for theory. For now, drag and layer-imbalance experiments of Liu at al. and Li et al. have gathered valuable information on these states.

Measurements of Liu et al. and Li et al. in double layer graphene have given a proof-of-principle demonstration of the intricate yet extensive structures that arise from the formation of two-component composite fermions. These studies have significantly expanded the phase diagram of the fractional quantum Hall effect and opened the door into other interesting phenomena. Re-entrant phase transitions as a function of the layer separation and imbalance, fractional quantum Hall states involving valley and spin degrees of freedom, interlayer correlated crystalline phases, effects of the Moiré patterns generated by the supporting layers, and new states supporting non-Abelian anyons are just a few of the exciting possibilities.